\documentclass[twocolumn,showpacs,aps]{revtex4}
\usepackage{psfig}
\def\la{\left\langle}
\def\ra{\right\rangle}
\begin{document}
\title{Universality of Velocity Gradients in Forced Burgers
  Turbulence}
\author{J\'er\'emie Bec} \email{bec@obs-nice.fr}
\affiliation{Observatoire de la C\^ote d'Azur, Lab.\ G.-D.\ Cassini,
  B.P.\ 4229, 06304 Nice Cedex 4, France}
\affiliation{CNLS -- Theoretical Division, LANL, Los Alamos, NM87545, USA}
\begin{abstract}
  It is demonstrated that Burgers turbulence subject to large-scale
  white-noise-in-time random forcing has a universal power-law tail with
  exponent -7/2 in the probability density function of negative velocity
  gradients, as predicted by E, Khanin, Mazel and Sinai (1997, Phys.\ Rev.\ 
  Lett.\ {\bf 78}, 1904). A particle and shock tracking numerical method gives
  about five decades of scaling.  Using a Lagrangian approach, the -7/2 law is
  related to the shape of the unstable manifold associated to the global
  minimizer.
\end{abstract}
\maketitle
  The universality of small-scale properties in fully developed
  Navier--Stokes (NS) turbulence is frequently investigated assuming
  that a steady state is maintained by a large-scale random force.
  For structure functions (moments of increments) universality with
  respect to the force is conjectured in the case of three-dimensional
  NS turbulence and proven for certain linear passive scalar models
  (see, e.g., Ref.~\cite{rmp}). The universality of probability
  density functions (p.d.f.) for velocity increments and gradients is
  a difficult question which, so far, has been mostly addressed within
  the framework of the pressureless model of Burgers turbulence,
  usually the one-dimensional Burgers equation
\begin{equation}
\partial_t u +u \partial_x u = \nu \partial_{xx} u + f(x,t)
\label{burg1d}
\end{equation}
with white-noise-in-time forcing \cite{cy95}. It is generally
conjectured that, when $\nu \to 0$ and the forcing is confined to
large scales, the tail of the p.d.f. of velocity gradients $\xi$ at
large negative values follows a universal power-law
$p(\xi)\propto|\xi|^{-\alpha}$. The actual value of the exponent is
however a matter of controversy. Let us briefly recall some of the
arguments found in the literature.

A standard approach is based on studying the inviscid limit of the
Fokker--Planck equation for the p.d.f.\ 
\begin{equation}
\partial_tp -\partial_\xi\left(\xi^2p\right) -\xi p  
+\nu \partial_\xi \left(\la\partial_{xx}\xi|\xi\ra p \right) = 
B\partial_{\xi\xi}p,
\label{fokker-planck}
\end{equation}
where the right-hand side expresses the diffusion of probability due
to the delta-correlation in time of the forcing. It was pointed out by
Polyakov~\cite{p95} that the inviscid limit of (\ref{fokker-planck})
contains anomalies due to the singular behavior of the dissipative
term $\nu \partial_\xi \left(\la\partial_{xx} \xi|\xi\ra p \right)$.
The value $\alpha =3$ is obtained if anomalies are ignored \cite{gk98}
or if a piecewise linear approximation is made for the solutions of
the Burgers equation \cite{bm96}.  An operator product expansion (OPE)
method borrowed from quantum field theory has been proposed for
evaluating such anomalies and an argument presented in favor of
$\alpha=5/2$ (actually, for velocity increments and infinite
systems)~\cite{p95}. However, this expansion leads to a relation
involving unknown coefficients which must be determined, e.g., from
numerical simulations~\cite{yc96b98}, and restricts the possible
values to $5/2 \leq \alpha \leq 3$~\cite{b01}. Anomalies cannot be
understood without a complete description of the singularities of the
solutions, such as shocks, and of their statistical properties. For
the case of a space-periodic system (as we shall assume), a crucial
observation made in Ref.~\cite{ekms97} is that large negative
gradients stem mainly from {\em preshocks}, that is the cubic-root
singularities in the velocity preceding the formation of
shocks~\cite{ff83}. A simple argument was given in Ref.~\cite{ekms97}
for determining the fraction of space-time where the velocity gradient
is less than some large negative value.  This leads to $\alpha=7/2$
provided preshocks do not cluster.  Determinations of the dissipative
anomaly of (\ref{fokker-planck}) have been made by formal matched
asymptotics~\cite{eve99} and by bounded variation
calculus~\cite{eve00}. With the assumption that shocks are born with
vanishing amplitude from isolated preshocks, the value $\alpha=7/2$
was obtained~\cite{eve99,eve00}. Other attempts to derive $\alpha=7/2$
using also isolated preshocks have been made~\cite{sevenhalf}. Note
that there are simpler instances, including time-periodic
forcing~\cite{bfk00} and decaying Burgers turbulence with smooth
random initial conditions~\cite{eve00,bf00}, which fall in the
universality class $\alpha=7/2$, as can be shown by systematic
asymptotic expansions using a Lagrangian approach. In the presence of
forcing, the key issues which remained to be settled are the possible
clustering of preshocks and, closely related to this, the possible
birth of shocks with non-vanishing amplitude. The results presented
hereafter almost completely rule out such possibilities.

Numerically solving the randomly forced Burgers equation in the limit of
vanishing viscosity in such a way as to obtain clean scaling for the p.d.f.\ 
of gradients represents a significant challenge. Broadly speaking, there are
two classes of methods. On the one hand, methods involving a small viscosity,
either introduced explicitly (e.g.\ in a spectral calculation) or stemming
from discretization (e.g.\ in a finite difference calculation). Viscosity
gives rise to a power-law range with exponent $-1$ at very large negative
gradients~\cite{gk98} whose presence will make the inviscid $|\xi|^{-\alpha}$
range appear shallower than it actually is, unless extremely high spatial
resolution is used. On the other hand, there are methods which directly
capture the inviscid limit with the appropriate shock conditions such as the
fast Legendre transform method of Ref.~\cite{nv94} (extended to the forced
case in Ref.~\cite{bfk00}). This method is very well adapted to decaying
Burgers turbulence with non-smooth Brownian-type initial data~\cite{vdun94}
but, with spatially smooth forcing, it leads to delicate interpolation
problems which have been overcome in the case of time-periodic
forcing~\cite{bfk00}; with white-noise-in-time forcing, it is difficult to
prevent spurious accumulations of preshocks leading to $\alpha=3$.  To avoid
such pitfalls, we develop a Lagrangian particle and shock tracking
method~\cite{notebennaim} which is able to cleanly separate smooth parts of
the solution and is particularly effective for identifying preshocks. The main
idea of the method is to consider the evolution of a set of $N$ massless point
particles accelerated by a discrete-in-time approximation of the forcing with
a uniform time step. When two of these particles intersect, they merge and
create a new type of particle, a shock, characterized by its velocity (half
sum of the right and left velocities of merging particles) and its amplitude.
The particle-like shocks then evolve as ordinary particles, capture further
intersecting particles and may merge with other shocks. In order not to run
out of particles too quickly, the initial small region where particles have
the least chance of being subsequently captured is determined by localization
of the global minimizer (see below).  The calculation is then restarted from
$t=0$ for the same realization of forcing but with a vastly increased number
of particles in that region. This method gives complete control over shocks
and preshocks \cite{notefilm} and allows an accurate determination of the
relevant statistical quantity while keeping a manageable number of degrees of
freedom.

\begin{figure}[h]
\centerline{\psfig{file=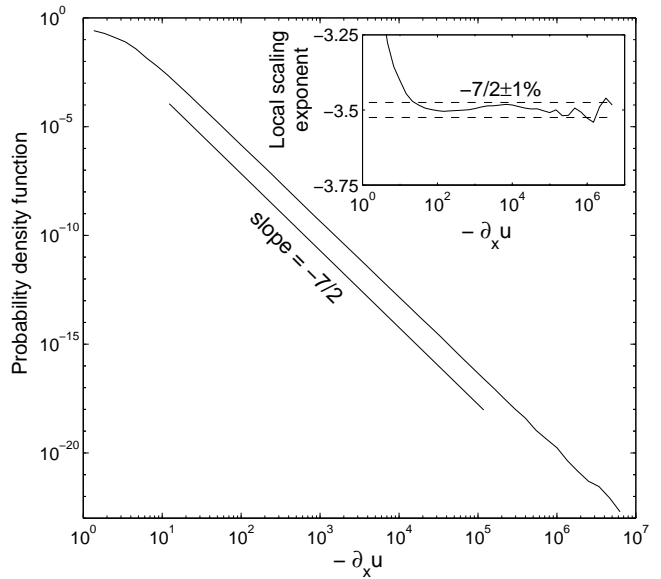,width=8.6cm}}
\caption{P.d.f.\ of the velocity gradient at negative values in log-log
  coordinates obtained by averaging over 20 realizations and a time
  interval of 5 units of time (after relaxation of transients). The
  simulation involves up to $N=10^5$ particles and the forcing is
  applied at discrete times separated by $\delta t=10^{-4}$.  Upper
  inset: local scaling exponent.}
\label{figpdf}
\end{figure}

Fig.~\ref{figpdf} shows the p.d.f.\ of the velocity gradients in
log-log coordinates at negative values for a Gaussian forcing
restricted to the first three Fourier modes with equal variances such
that the large-scale turnover time is order unity. Quantitative
information about the value of the exponent is obtained by measuring
the ``local scaling exponent'', i.e.\ the logarithmic derivative of
the p.d.f.\ calculated here using least-square fits on half-decades.
It is seen that over about five decades, the local exponent is within
less than 1\% of the value $-7/2$ predicted by E {\it et
  al.}~\cite{ekms97}. This value of the exponent was also obtained
numerically (with a fewer particles) for other large-scale forcing
instances with compactly supported or exponentially decreasing spectra
and also for non-Gaussian forcing (e.g.\ with Fourier amplitudes
having a Bernoulli distribution or an uniform distribution in an
interval). Evidence for non-clustering of preshocks is obtained by
counting the average number of shock formations per unit time.  For
different types of large-scale forcing, we found that the typical mean
number of preshocks per turnover time is comparable to the number of
forcing Fourier modes significantly excited.  For such forcings, the
density of preshocks is found to vary by not more than 6\% when the
time step varies by two orders of magnitude around $\delta t=10^{-4}$,
which is hardly consistent with a power-law (and even a logarithmic)
divergence as $\delta t\to 0$. Furthermore, we have checked that
shocks are always born with vanishing amplitude (within numerical
errors).

\begin{figure}[h]
\centerline{\psfig{file=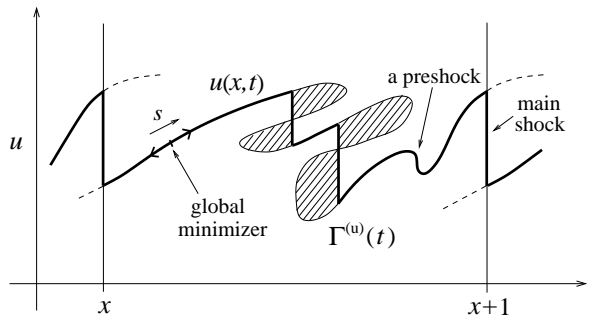,width=8cm}}
\caption{Sketch of the unstable manifold $\Gamma^{\rm (u)}$ in the
  $(x,u)$ plane at a time $t$ with a preshock occurring. Shock
  locations are obtained by applying Maxwell rules to the loops. The
  velocity is shown as a bold line.}
\label{figvinstable}
\end{figure}

Turning now to theoretical results, let us briefly recall the construction of
solutions developed by E {\it et al.}~\cite{ekms00}, in terms of the dynamical
system associated to the characteristics of (\ref{burg1d}) in the inviscid
limit \cite{note}. The force is assumed to derive from a Gaussian potential
$F(x,t)$, delta-correlated in time, periodic of period 1 and analytic in
space. A statistically stationary r\'egime is reached by taking the initial
time at $-\infty$. The central point of the construction is the following
variational characterization of the solution at an arbitrary time ($t=0$
chosen for convenience):
\begin{equation}
u(x,0) = {\partial \over \partial x} \min_{X(\cdot)} \left[ \int_{-\infty}^{0}
  \left[ {1\over2}\dot{X}^2(t) -F\left(X(t),t\right) \right] dt  \right],
\label{minimum}
\end{equation}
where the minimum is taken over all piecewise smooth (absolutely continuous)
curves $X(t)$ with $t\in (-\infty,0]$ such that $X(0)=x$. A curve minimizing
the action in~(\ref{minimum}) is called a {\em minimizer} and should be
understood as a fluid particle trajectory. It obviously has to satisfy for all
$t<0$ the Euler--Lagrange equations:
\begin{eqnarray}
\dot{X}(t) &=& U(t), \label{eulagX} \\
\dot{U}(t) &=& f(X(t),t). \label{eulagU}
\end{eqnarray}
Except for a finite number of $x$-values, there exists a unique
minimizer~\cite{ekms00}. The locations where there are more than one minimizer
correspond to shocks. The minimizers converge exponentially fast backward in
time to the trajectory of the unique fluid particle which is never absorbed by
a shock.  This trajectory is called the {\em global minimizer} because its
action is minimal at any time; it corresponds to a hyperbolic trajectory of
the dynamical system (\ref{eulagX})-(\ref{eulagU}).  Associated to it, there
are two curves in the phase-space $(x,u)$: a stable (attracting) manifold
$\Gamma^{\rm (s)}$ and an unstable (repulsive) manifold $\Gamma^{\rm (u)}$.
The minimizers converge backward in time to the global minimizer and, thus,
the graph of the solution is made of pieces of the unstable manifold with
jumps at shocks. One of these shocks, called the {\em main shock}, is singled
out. It is the unique shock which has always existed in the past (whereas
generic shocks are born at some finite time $t<0$); it may be shown that it
corresponds to the position giving rise to the left-most and the right-most
minimizers which approach the global one backward in time.  The other shocks
cut through the doublefold loops of the unstable manifold (see
Fig.~\ref{figvinstable}). We observe that their locations can be obtained by a
Maxwell rule applied to those loops. Indeed, the difference of the two areas
defined by cutting such a loop at some position $x$ is equal to the difference
of actions of the two minimizers defined by the upper and lower branches and,
thus, vanishes at the shock location.

We also observe that the structure just outlined has much in common with that
appearing in the {\em unforced\/} Burgers equation. Indeed when $f=0$, the
solution to the Burgers equation can be constructed from the Lagrangian
manifold in the $(x,u)$ plane, defined as the position and the velocity of
fluid particles when ignoring shocks. This manifold is parameterized by the
Lagrangian coordinate $a$; denoting $u_0$ the initial velocity, we then have
simply $x=a+tu_0(a)$ and $u=u_0(a)$. The actual solution with shocks is
obtained by applying the standard Maxwell rule to the Lagrangian manifold. In
the forced case, a parameterization of the unstable manifold (e.g.\ by the
arclength) is now the analog of the Lagrangian coordinate. But there are two
important differences: first, in the unforced case, the time evolution of the
Lagrangian manifold is explicit and linear while, in the presence of a force,
the Euler--Lagrange equations (\ref{eulagX})-(\ref{eulagU}) are not, in
general, explicitly solvable and the unstable manifold has a hyperbolic
dynamic. Second, the smoothness of the Lagrangian manifold in the unforced
case stems directly from the smoothness of the initial data, whereas in the
forced case Pesin's theory must be used to show that when the force is
indefinitely differentiable in space, so is the unstable
manifold~\cite{ekms00}.

Using the smoothness of the unstable manifold, we now formally derive
the $-7/2$ law, by an argument mostly borrowed from the unforced
case~\cite{bf00}. Let $\Gamma^{\rm (u)} = \{ (X(s),U(s))\}$ with $s$
real, be a parameterization of the unstable manifold at time $t=0$. It
is assumed for convenience that $s=0$ corresponds to the global
minimizer and that $X'(0)>0$, where primes denote $s$-derivatives. The
velocity is exactly obtained by eliminating from the unstable manifold
the shaded areas determined by the Maxwell rules and the parts beyond
the main shock (shown as dashed lines in Fig.~\ref{figvinstable}). The
surviving set of parameter values (excluding shocks) is denoted
$\Omega$. Turning to the statistical description, the p.d.f.\ of
velocity gradients may be written
\begin{equation}
p(\xi)=\la \delta\left (\partial_x u(x,0)-\xi\right)\ra.
\label{pdf1}
\end{equation}
Because of homogeneity, we can integrate over the space period and then change
from the $x$ variable to the $s$ variable to obtain
\begin{equation}
p(\xi)=\int_0^1 p(\xi) dx
=\la \int_\Omega \delta\left ({U' \over X'}-\xi\right) X' ds\ra.
\label{pdf2}
\end{equation}
Note that since a finite gradient is assumed, $x$ cannot be at a shock
position.  Denoting by $s_k$ the parameter values where the argument of the
delta function vanishes, we obtain
\begin{equation}
p(\xi)=\la \sum_k {X'^2 \over |U'' -\xi X''|} \delta(s-s_k)ds\ra.
\label{pdf3}
\end{equation}
For very large negative values of $\xi$, the $s_k$'s must be near some
$s_{*j}$, corresponding to a local minimum of $X'$. Taylor expansions of $X$
and $U$ in the vicinities of the $s_{*j}$'s and the use of the Maxwell rule
show that the $s_{*j}$'s are located in space-time near preshocks satisfying
$X'=0$ and $X''=0$ with $X'''>0$ (see Fig.~\ref{figvinstable}). Proceeding as
in Ref.~\cite{bf00}, we finally obtain, to leading order
\begin{eqnarray}
&p(\xi)\simeq C |\xi|^{-7/2}, \qquad \xi\to -\infty,\\
\label{pdf4}
&C = {5\sqrt{2}\over2} \la \int_{\Omega, X'''>0} |X'''|^{1/2} |U'|^{5/2}
\delta(X')\delta(X'') ds\ra.
\label{cte}
\end{eqnarray}
Hence, the constant involves the mean of $|X'''|^{1/2} |U'|^{5/2}$ at
preshocks. Its evaluation requires the knowledge of the joint probability
distribution of $X'$, $X''$, $X'''$ and $U'$. From the Euler--Lagrange
equations (\ref{eulagX})-(\ref{eulagU}), we observe that a set of ordinary
differential equations with nonlinear stochastic forcing is easily obtained
for $X$, $U$ and the aforementioned four variables. From these equations,
using techniques similar to those developed in Ref.~\cite{ekms00} (where a
subset of these stochastic equations is studied), it should be possible, on
the one hand, to make our derivation more rigorous (including for the
non-clustering of preshocks) and, on the other hand, to obtain an upper bound
for the constant $C$ in the $-7/2$ law. Note that the expression for $C$
involves also an integral over the admissible set of parameters $\Omega$ whose
determination cannot in general be done by local analysis with ordinary
differential equations. This is why only an upper bound is expected.

As noted in Ref.~\cite{ekms97}, the universality with respect to the forcing
of the p.d.f.\ of large negative velocity gradients may be extended to
negative velocity increments, provided that they are not significantly
influenced by shocks. Without understanding of all the mechanisms leading to
small-amplitude shocks in the forced case, the issue of universality for the
p.d.f.'s of velocity increments cannot be settled. A first step would be to
determine numerically the distribution of shock amplitudes. Note that our
technique may also be extended to the case of forcing at scales much smaller
than the size of the system, a problem close to that considered by
Polyakov~\cite{p95}, which is left for future work.

I wish to express my gratitude to U.~Frisch and K.~Khanin for numerous
helpful discussions and encouragements. I also thank M.~Blank,
G.~Eyink, A.~Fouxon, R.~Mohayaee, E.~Vanden Eijnden, M.~Vergassola and
V.~Yakhot for useful remarks.  Work in Nice was supported by the
European Union, under contract HPRN-CT-2000-00162; work at CNLS was
supported by the U.S.\ Department of Energy, under contract
W-7405-ENG-36. Simulations were performed on Avalon at the Los Alamos
National Laboratory.


\begin{thebibliography}{99}
\bibitem{note}
  An  elementary introduction to the concepts of
  minimizers and main shock  may be found in Ref.~\cite{bfk00} for the case of
  periodic forcing.
\end{thebibliography}
\end{document}